\documentstyle[prl,psfig,aps,epic,eepic,epsfig,graphics,rotating]{revtex}
\draft
\begin{document}
\title{Observation of Proximity Resonances in a Parallel-Plate Waveguide}
\author{
  J. S. Hersch$^{1,*}$ and  E. J. Heller$^{1,2,\dagger}$\\
  }
\address{
  $^1$Department of Physics, Harvard University, Cambridge,
  Massachusetts 02138\\ 
  $^2$Harvard-Smithsonian Center for Astrophysics, Cambridge,
  Massachusetts 02138\\ 
  $^*$e.mail: hersch@monsoon.harvard.edu\\
  $^\dagger$e.mail: heller@physics.harvard.edu
  }
\maketitle
\begin{abstract}
  Experiments with dielectric scatterers in a parallel-plate 
  waveguide have
  verified for the first time the existence of proximity resonances in 
  two dimensions.  A numerical solution to the scattering problem supports
  the analysis of the experimental data.
\end{abstract}

  It has recently been shown that two resonant s-wave
  scatterers placed close together produce two
  resonances in the spectrum of the combined system \cite{rick}.
  The first, which remains s-wave in character, is shifted down
  in energy and broadened with respect to the original single scatterer
  resonance. The second resonance, which is p-wave in character,  is shifted
  up an equal amount in energy and can have a very narrow width. In fact,
  the width of the p-wave resonance
  vanishes as the scatterers approach each other.
  This second resonance has been dubbed the proximity resonance.
  
  Proximity resonances are important in a number of physical contexts,
  including scattering of sound from small identical bubbles in
  liquids \cite{bubble1,bubble2},
  and scattering and emission of light from nearby dipole scatterers
  \cite{devoe,berman}
  where a proximity resonance  effect has long been known under the 
  name of Dicke super-radiance and sub-radiance.  In Ref. \cite{rick},
  the effect was discussed for particle scattering from two identical 
  atoms (or other identical scatterers) for the first time.  Here we discuss
  yet another context, the classical scattering of electromagnetic waves
  from dielectric discs.  At the same time (however see the caveat below)
  the system we describe mimics quantum scattering from two adjacent
  potential wells
  in two dimensions\cite{Sridhar,stock}.
  
  For the purposes of modeling the
  experiment, we developed a method of solving the scattering problem 
  involving cylindrical basis functions centered on each disc.  It turned
  out that
  the point scatterer model \cite{rodberg,drukarev}, 
  which was used in the original discussion
  of proximity resonances \cite{rick}, was not sufficient to accurately
  model the
  experiment. In order for the point scatterer model to be applicable, at
  least two conditions must be met:  $r \ll \lambda$, and $r \ll d$,
  where $r$ is the physical radius of each scatterer, $\lambda$ the
  wavelength, and $d$ the distance between the scatterers.  In our
  experiments, the first condition was always met, but the second
  was not. 
  
  Other work \cite{kristi} indicates that there may be a
  similar effect present in the bound state spectrum of two nearby
  dielectric discs in a parallel-plate waveguide. Szmytkowski {\it et al}.
  \cite{polish} have found theoretically a similar resonance with fixed
  scattering length point interactions.
  
  The picture to keep in mind when thinking about the proximity resonance is
  the following: imagine two nearby point sources of unit amplitude,
  situated much closer together than a wavelength.  When these sources are
  in phase, amplitude will add up nearly in phase everywhere in space, and
  the amplitude far from the sources will be appreciable. The far
  field intensity clearly will be s-wave in character.   When the sources
  are out of phase, amplitude will interfere destructively everywhere, and the
  far field intensity will be much reduced compared to the
  in-phase case.  Now, for a scattering resonance, the width of
  the resonance is proportional to the rate at which amplitude escapes
  from the neighborhood of the scattering system.  This rate is
  proportional to the ratio
  $|\psi_{\text{far}}/\psi_{\text{near}}|^2$, where $\psi_{\text{far}}$ is
  the far field amplitude and $\psi_{\text{near}}$ is the
  amplitude in the near field. This ratio will remain finite for the
  in-phase
  pair of scatterers, and vanish for the out of phase pair, as their
  separation goes to zero.  This narrows the proximity resonance as the
  scatterers
  are brought closer together.

\begin{figure}
  \epsfig{figure=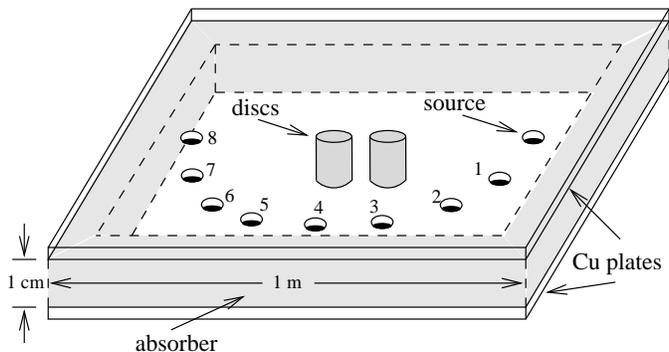}
  \caption{\label{cavity} The scattering arena.  Source and receiving
    antennae were inserted through holes drilled in the top plate.
    The field could be measured in any of eight locations located 
    on a semicircle 25 cm from the discs.  Note that
    the figure is not drawn to scale.}
\end{figure}

  The waveguide, shown in Fig.~\ref{cavity}, consisted of
  two parallel copper plates, 1 m square,
  separated by a 1 cm gap.  To minimize the effect of waves reflected
  off the edges of the waveguide, the
  perimeter was lined with a 11.5 cm thick layer of microwave
  absorber (C-RAM LF-79, Cuming Microwave Corp.), designed
  to provide 20 dB of attenuation in the reflected wave intensity at
  frequencies above 600 MHz.  
  Without the absorber, there would be substantial
  reflections of both the incident and scattered wave off the edges of
  the waveguide, which would produce strong cavity modes and
  unnecessarily complicate the analysis. 
  The important effect of the absorber was to allow the
  waveguide to behave as if it were infinite in extent in the directions
  parallel to the plates, and thus support oscillations at all frequencies.
    
  The scatterers were cylindrical in shape (radius: 2 mm, height: 1
  cm) and had a measured dielectric constant of $\epsilon = 77\pm1$. 
  Each disc had an individual
  s-wave scattering resonance at 2.3 GHz with a 1.1 GHz width. They were
  illuminated with microwaves from a point source located 25 cm away from
  the midpoint of the two scatterers.  The field in the waveguide could be
  measured at eight points located on a circle of 25 cm radius centered at
  the midpoint between the scatterers. 
  
  Antennae were inserted perpendicular to the plates to launch the
  incident wave and measure the field.  Such antennae couple to an electric
  field perpendicular to the plates.  For a plate separation of 1 cm
  and frequencies below 15 GHz (the experiment operated between 1-3 GHz),
  {\em only} the TEM (transverse electromagnetic) mode propagates in 
  the waveguide, and all others are
  evanescent. The classification TEM means that, for this mode, both
  $\vec E$ and $\vec H$ are transverse to the direction of
  propogation. In particular, $\vec E$ is everywhere perpendicular to
  the plates, and $\vec H$ is everywhere parallel to the plates.
  As an example, we calculate the decay constant, \mbox{$\kappa =
  \sqrt{k_z^2-(\omega/c)^2}$}, for the mode with one oscillation
  transverse to the plates at 3 GHz.  With $k_z = \pi/L$ and $L = 1$
  cm, we find $\kappa \simeq 3 \ \mbox{cm}^{-1}$. This means that this
  mode has decayed by a factor $e^{-75}$ over a distance of 25 cm, the
  distance between the source and the scatterers.
  Thus we may safely ignore all modes but the TEM mode for the purpose
  of this work.

  As mentioned above, for the TEM mode both $\vec E$ and $\vec H$ 
  are transverse to the 
  direction of  propagation, just as for a plane wave in free space. In fact, a
  useful visualization of this mode in the waveguide is just a section of an
  infinite plane wave, $\vec E_0 e^{i \vec k \cdot \vec r}$, with wave 
  vector $\vec k$ parallel, and electric field $\vec E_0$  normal to the
  plates.  Furthermore, the TEM mode has {\em no variation} of the fields in
  the direction perpendicular to the plates and is thus truly two
  dimensional\cite{sridhar1}. 
  In fact, the entire field structure $\{\vec E(x,y),\vec H(x,y)\}$ can  
  be derived from knowledge of $E_z(x,y)$ alone \cite{jackson}, where $z$ is
  understood to be the direction perpendicular to the plates.
  Furthermore, for the TEM mode, the
  boundary conditions on $E_z$ at the dielectric surface are identical
  to those of a quantum mechanical square well: $E_z$ and its normal
  derivative, $\partial_n E_z$, must be continuous across the
  interface. Thus the component $E_z$ in the waveguide plays
  the role of $\psi$ in a two dimensional quantum mechanical
  system. Henceforth we will refer to $E_z$ as $\psi$.
  
  However, there remains one important difference between dielectrics 
  and quantum mechanical square wells. In quantum mechanics, the
  ratio of wavenumbers inside and outside the well is
  \[
  \frac{k_{\text{in}}}{k_{\text{out}}} = \sqrt{\frac{E-V}{E}},
  \]
  where $V$ is the well depth and $E$ is the energy.  Note that this
  ratio depends on $E$, and diverges at low energy.  In the
  electromagnetic case, this ratio is a constant, and equal to the
  index of refraction:
  \[
  \frac{k_{\text{in}}}{k_{\text{out}}} = \sqrt{\epsilon}.
  \]
  This means that a system of quantum square wells can only be compared
  with an equivalent system of dielectric discs at a particular energy.
  If the energy is changed, $\epsilon$ must also be changed to retain
  correspondence.
  
  The measured signal was compared
  to the source signal in both amplitude and phase with a HP 8714C network
  analyzer.  Because both amplitude and phase could be measured,
  it was possible to extract the (complex) scattered wave, $\psi_s$ from the
  full
  signal, $\psi = \psi_0 + \psi_s$, where $\psi_0$ is the incident
  wave. This was done by removing the
  scatterers from the waveguide and repeating the measurement, yielding
  $\psi_0$.  
  This result was then subtracted from the full wave to yield the
  scattered wave signal.

\begin{figure}
  \epsfig{figure=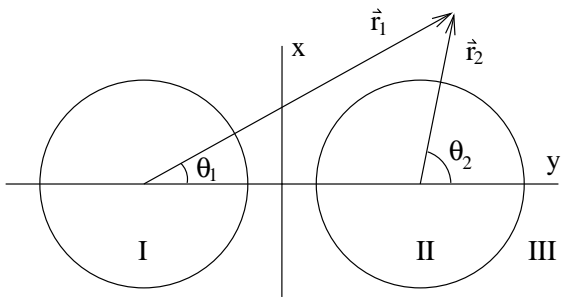}
  \caption{\label{discs} A coordinate system for two disc scattering.}
\end{figure}

  The full solution to the scattering problem of a single dielectric
  disc in a parallel plate waveguide can be found analytically \cite{vdH}.
  The two disc problem, however, becomes difficult because of the lack 
  of cylindrical symmetry.  We address this difficulty by using a basis
  which reflects the broken symmetry of the problem:  two sets of Bessel
  functions, each centered on one of the discs. This method is similar
  in spirit to that of Goell \cite{waveguide}. Referring to
  Fig. \ref{discs}, we have in regions I, II, and III, respectively,
  \[
  \psi_I = \sum_{l=-l_{\text{max}}}^{l_{\text{max}}} A_l J_l(\kappa r_1)
  e^{i l \theta_1},
  \]
  \[
  \psi_{I\!I} = \sum_{l=-l_{\text{max}}}^{l_{\text{max}}} B_l J_l(\kappa
  r_2) e^{i l \theta_2},
  \]
  and
  \[
  \psi_{I\!I\!I} = \psi_0 + \sum_{l=-l_{\text{max}}}^{l_{\text{max}}}
  \left[C_l H_l^{(1)}(k r_1) e^{i l \theta_1} + D_l H_l^{(1)}(k r_2) 
    e^{i l \theta_2}\right],
  \]
  where $J_l(x)$ and $H_l^{(1)}(x)$ are Bessel functions and
  Hankel functions of the first kind,
  $\psi_0$ is a TEM incident wave, $\kappa  = \sqrt{\epsilon} \ k$,
  and  $l_{\text{max}}$ determines the size
  of the basis set. Note that the variable $z$ does not appear in the
  above equations, because for the TEM mode there is no $z$
  dependence of the fields.  An exact solution would require that 
  $l_{\text{max}} \to \infty$.  However, we find very good solutions  
  for $l_{\text{max}}$ as
  small as 5.  The complex constants $A_l,B_l,C_l,D_l$ are to be
  determined by matching $\psi$ and its normal derivative $\partial_n
  \psi$ along the perimeter of each disc.  

  The exact solution would require matching $\{\psi,\partial_n
  \psi\}$ at all points along the boundary of each disc.  In practice,
  one can only match at a finite number of points.  From each matching point,
  one obtains two equations relating the constants $A_l,B_l,C_l,D_l$.
  The entire collection of matching equations can be expressed in
  matrix form, $Mx=b$,
  where the number of rows and columns of $M$ is determined by the number of
  matching points and basis functions, respectively.  The vector $x$ is
  built up of the coefficients $A_l,B_l,C_l,D_l$, and $b$ is determined
  by the incident wave $\psi_0$. In general, one chooses more matching
  points than basis functions, so that the solution $x$ minimizes the
  length $r = |Mx - b|$.  This minimization is efficiently carried out
  by finding the singular value decomposition of the matrix $M$ \cite{NR}.
  The residual $r$ provides an indication of
  the accuracy of the solution.  For this work, typical values or $r$
  were $10^{-10}$ per matching point.  This is to be compared with
  values of $|\psi|$ and $|\partial_n \psi|$ of order unity on the
  perimeters of the discs. 

\begin{figure}
  \epsfig{figure=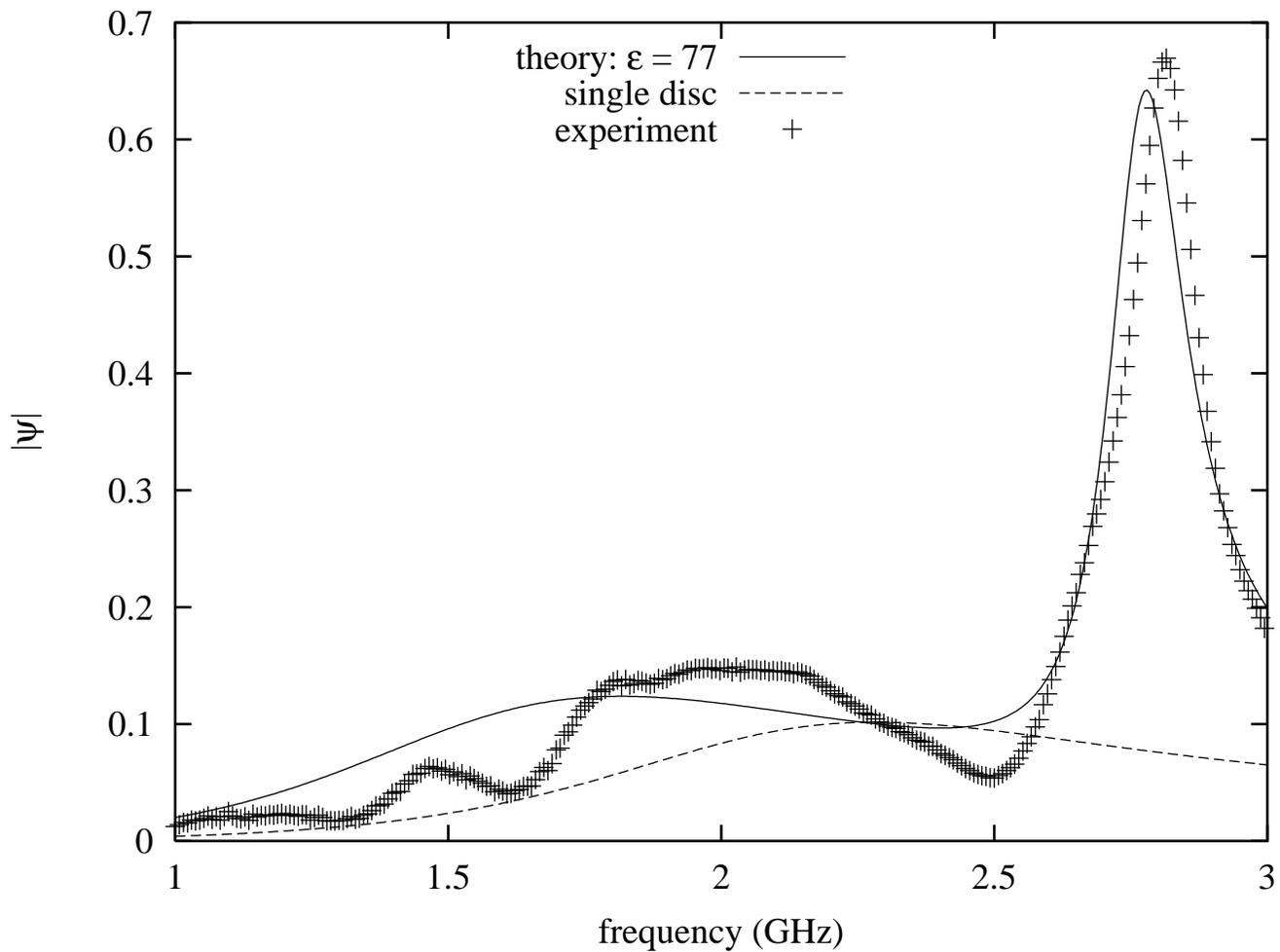}
  \caption{\label{amp} Here we plot the scattered amplitude at
    position 7 versus frequency. 
    Comparison between theoretical (solid line) and
    experimental data (crosses).  Disc separation: 1.0 cm.  
    The single disc resonance also shown (dashed line).}
\end{figure}

  In Fig. \ref{amp} we plot the scattered amplitude $|\psi_s| =
  |\psi - \psi_0|$ measured at position 7 (see Fig. \ref{cavity}).
  The theoretical result agrees very well, apart from a 
  weak 0.3 GHz modulation of the experimental signal due to reflections
  off the absorbing walls of the waveguide. The
  numerical data was generated using $l_{\text{max}} = 5$ and matching at 10
  equally spaced locations around each disc.     The broad feature centered
  around 2.0 GHz is the s-wave (in phase) resonance.  A strong
  proximity resonance is apparent at around 2.8 GHz. The width of this
  peak is smaller by a factor of 7 than the single
  scatterer s-wave resonance width.  We also checked that this peak
  was indeed p-wave in character by measuring the angular dependence of
  the scattered wave in the vicinity of 2.8 GHz.  Notably, the peak was
  absent when the measuring antenna was placed on the line equidistant
  from each disc, which defines a nodal line of the scattered wave for
  a p-wave resonance. 
  
\begin{figure}
  \epsfig{figure=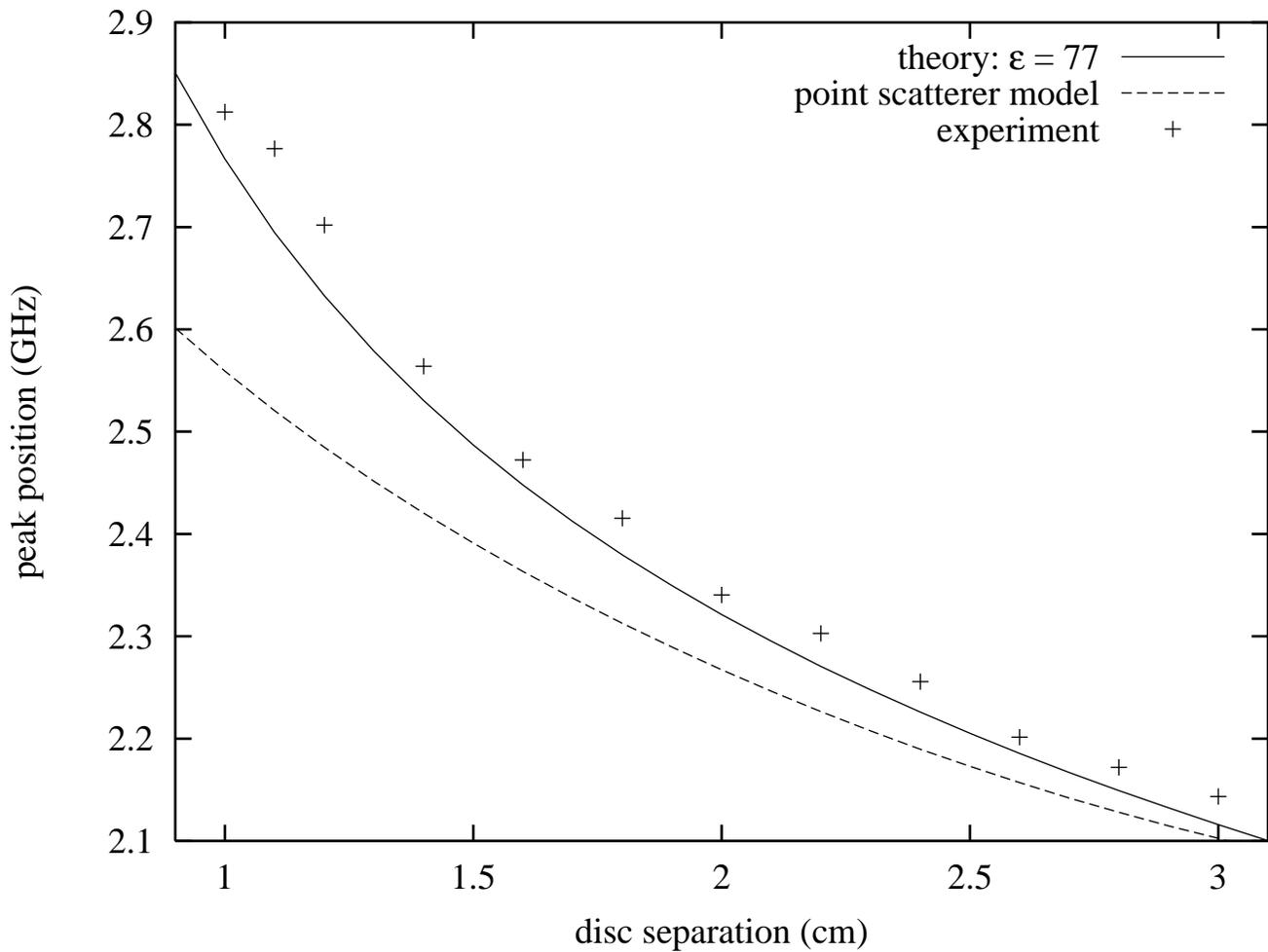}
  \caption{\label{peak} Here we plot the position of the proximity
    resonance peak versus disc separation. The theoretical curve
    (solid line) tracks the experimental values well (crosses).  The
    point scatterer model prediction is also shown (dashed line). }
\end{figure}

\begin{figure}
  \epsfig{figure=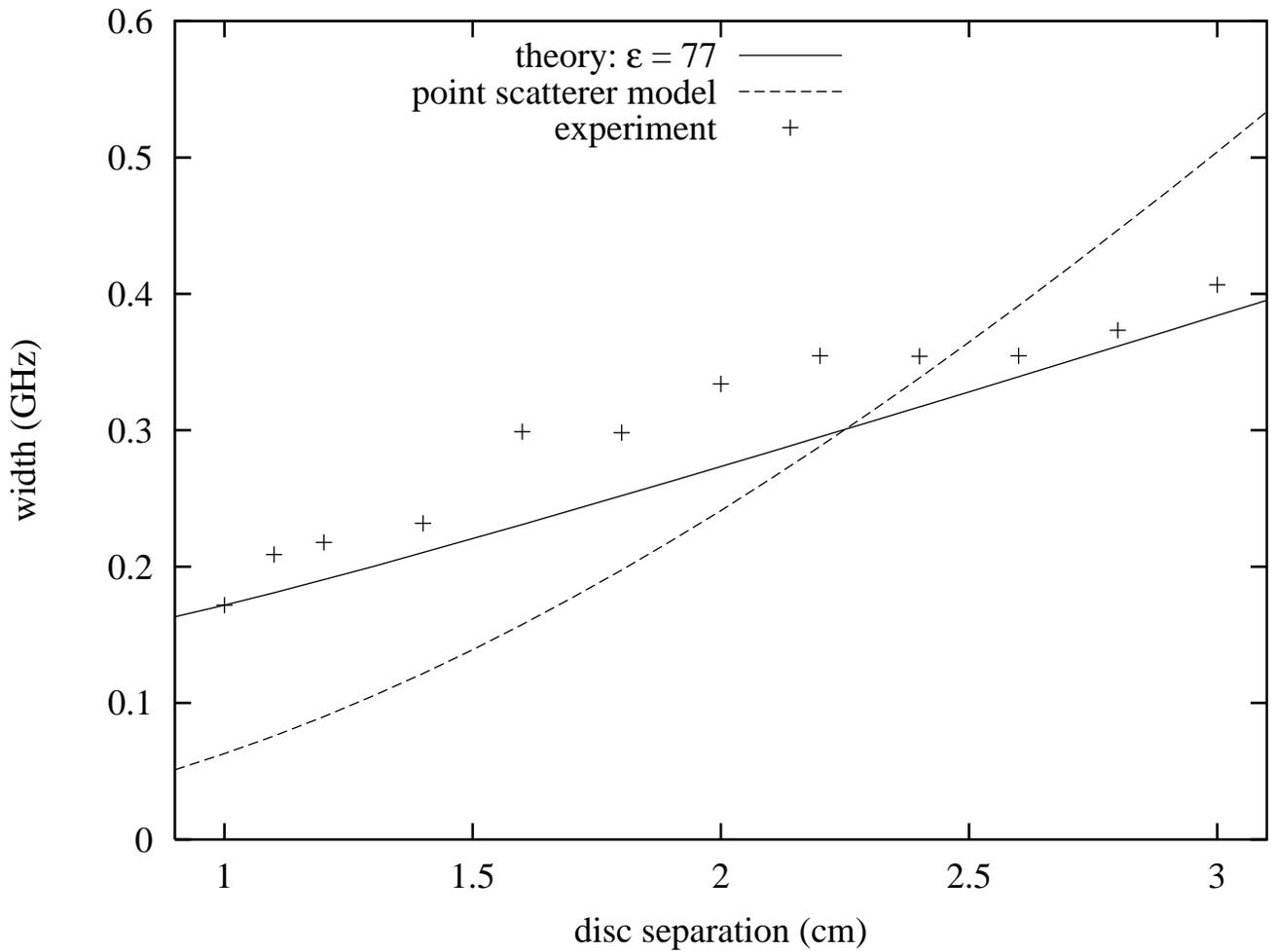}
  \caption{\label{width} Here we plot the width of the proximity
    resonance versus disc separation. As above, the theoretical curve
    (solid line) models the experimental data (crosses) well.  The
    point scatterer model prediction is also shown (dashed line).}
\end{figure}

  In Figs. \ref{peak}, \ref{width} we plot the peak position and width,
  respectively, of the proximity resonance as a function of disc
  separation.  Again the numerical predictions are in good agreement
  with the data.
  For comparison, we also include the predictions of
  the cruder point scatterer model, using as input parameters a single
  scatterer resonance frequency $f_0 = 2.3$ GHz and width of $\Gamma_0 =
  1.1$ GHz. 
  It can be shown that within this model, the
  proximity resonance peak position $f$ and width $\Gamma$ obey the
  following formulae,
  \[
  f = f_0 - \frac{\Gamma_0}{2}Y_0(k_0 d)\ ,\ \ \Gamma =
  \Gamma_0\left(1-J_0(k_0 d)\right),
  \]
  where $J_0$ is a zeroth order Bessel function, $Y_0$ is a zeroth
  order Neumann function,
  $k_0$ is the on-resonance wave number of a single scatterer, and $d$
  is the distance between the scatterers.  The
  point scatterer model does a good job of tracking the peak positions,
  but the resonance widths are not described well by the model.
  
  In summary, we have, for the first time, observed proximity
  resonances in a two dimensional system.  The analysis of the 
  experimental data seems
  to be well supported by a numerical solution to the scattering
  problem.  Immediate extensions of the ideas presented here include
  increasing the number of scatterers to look for even narrower
  resonances, which would be associated with higher angular momentum
  scattered waves (d-waves, for example).  The spectrum of a dense
  (compared to a wavelength), ordered array of s-wave resonant
  scatterers is also an interesting system as it relates to band
  structure formation.  

  We acknowledge   S. Sridhar and K. Pance
  for suggesting the use of high $\epsilon$ dielectrics
  in our experiment.
  We would like to thank P. Horowitz for advice on microwave
  techniques, and especially for the use of his network analyzer.  J.
  Doyle provided insight and encouragement.  This work was supported 
  through funding from  Harvard University, and the National Science
  Foundation, through ITAMP 
  and also Grant NSF-CHE9610501.

\end{document}